\documentclass[%
 reprint,
 amsmath,amssymb,
aps,
prb,
superscriptaddress
]{revtex4-2}

\usepackage{color}
\usepackage{graphicx}
\usepackage{dcolumn}
\usepackage{bm}

\usepackage{float}

\begin{document}

\preprint{APS/123-QED}

\title{Exceptional point induced unidirectional radiation from non-Hermitian plasmonic structures}

\author{Yuto Moritake}
\altaffiliation{moritake@phys.titech.ac.jp}
\affiliation{Department of Physics, Tokyo Institute of Technology, 2-12-1 Ookayama, Meguro-ku, Tokyo 152-8550, Japan}
\affiliation{PRESTO, Japan Science and Technology Agency, 4-1-8 Honcho, Kawaguchi, Saitama 332-0012, Japan}

\author{Masaya Notomi}
\affiliation{Department of Physics, Tokyo Institute of Technology, 2-12-1 Ookayama, Meguro-ku, Tokyo 152-8550, Japan}
\affiliation{NTT Basic Research Laboratories, NTT Corporation, 3-1 Morinosato-Wakamiya, Atsugi-shi, Kanagawa 243-0198, Japan}
\affiliation{Nanophotonics Center, NTT Corporation, 3-1, Morinosato-Wakamiya, Atsugi-shi, Kanagawa 243-0198, Japan}

\date{\today}

\begin{abstract}
Non-Hermitian (NH) photonics has attracted considerable attention from researchers owing to exotic properties that originate from the parity-time ($\mathcal{PT}$) phase transition and exceptional points (EPs).
To date, the $\mathcal{PT}$ phase transition, EPs, and circling around EPs have been investigated in many photonic systems.
However, few studies focused on the singular nature of the EP eigenstates of the Hamiltonian matrices.
Moreover, the switching of an EP eigenstate based on the sign of the coupling constant and its manifestation in physical phenomena have not yet been investigated.
In this paper, we propose and numerically demonstrate a unidirectional radiation phenomena manifested by the formation of Huygens dipoles using the singular EP eigenstates in coupled plasmonic systems.
Two types of EPs corresponding to positive and negative signs of the coupling constants are realized using dipole-dipole coupling.
We show that the Huygens dipole is formed at the EP condition and its radiation direction can be controlled by choosing the sign of the coupling.
The unique photonic functionality, which originates directly from the singular eigenstate at the EP uncovers a new aspect of NH photonic physics and devices.
\end{abstract}

\maketitle

\section{\label{sec:level1}Introduction}
Physical phenomena in systems described by non-Hermitian (NH) Hamiltonians have attracted considerable attention from researchers.
The eigenvalues of NH Hamiltonians are generally complex.
However, in 1998, Bender and Boettcher showed that real eigenvalues are allowed in NH systems if Hamiltonians have parity-time ($\mathcal{PT}$) symmetry \cite{Bender1998}.
The systems described by NH Hamiltonian with $\mathcal{PT}$ symmetry exhibit the $\mathcal{PT}$ phase transitions from the $\mathcal{PT}$ symmetric phase to the $\mathcal{PT}$ broken phase across exceptional points (EPs).
The eigenvalues of the systems are real or purely imaginary in the $\mathcal{PT}$ symmetric or broken phases, respectively.
More interestingly, the eigenvalues and eigenmodes coalesce at the EP, which is a unique property of NH systems with $\mathcal{PT}$ symmetry \cite{Heiss2012}.

In photonics, non-Hermiticity can be easily implemented through radiation loss and material gain/loss.
Therefore, photonic systems are feasible platforms for examining and utilizing NH physics in realistic devices \cite{Feng2017}.
So far, many studies that demonstrated a variety of intriguing phenomena in NH photonics have been reported.
These phenomena include asymmetric oscillation \cite{Ruter2010}, nonlinearity-based non-reciprocal transmission \cite{Peng2014}, reverse pump dependence \cite{Brandstetter2014}, enhancement of sensitivity \cite{Hodaei2017,Park2018}, asymmetric propagation \cite{Feng2013,Feng2014,Huang2017}, a superluminal effect \cite{Takata2017}, encircling around EPs \cite{Doppler2016,Yoon2018,Hassan2017}, large local density of states \cite{Takata2021}, coherent perfect absorption \cite{Sun2014,Wong2016}, and loss-induced transparency \cite{Guo2009}.

A typical example of photonic systems described by NH Hamiltonian with $\mathcal{PT}$ symmetry is a doubly coupled resonator, which is described by the following 2 $\times$ 2 matrix:
\begin{equation}
\begin{split}
\label{Hamiltonian}
    \cal{H} &=
    \left( \begin{array}{cc}
    \omega_{1} +i \gamma_{1} & \kappa \\
    \kappa & \omega_{2} +i \gamma_{2}
    \end{array} \right)
    \\&=
    \left( \begin{array}{cc}
    \bar{\omega} +i \bar{\gamma} & 0 \\
    0 & \bar{\omega} +i \bar{\gamma}
    \end{array} \right)
    \\&+
    \left( \begin{array}{cc}
    \Delta \omega +i \Delta \gamma & \kappa \\
    \kappa & -\Delta \omega -i \Delta \gamma
    \end{array} \right)
\end{split}
\end{equation}
Here, $\omega_{i}$ and $\gamma_{i}$ ($i = 1, 2$) are the resonant frequency and loss/gain rate of the resonators, respectively.
$\kappa$ is a coupling constant and is generally complex: $\kappa=\kappa^{\prime}+i\kappa^{\prime\prime}$.
$\bar{\omega}$ and $\bar{\gamma}$ are the averages of $\omega_{i}$ and $\gamma_{i}$, respectively.
$\Delta \omega$ and $\Delta \gamma$ are the difference between $\omega_{i}$ and $\gamma_{i}$, respectively.
If $\bar{\gamma}=\Delta\omega=\kappa^{\prime\prime}=0$, Hamiltonian in Eq. \ref{Hamiltonian} has $\mathcal{PT}$ symmetry.
However, even if $\bar{\gamma}$ has a finite value, it is known that the system exhibits NH physics such as the $\mathcal{PT}$ phase transition and EP.
Such systems are called loss (or gain)- biased $\mathcal{PT}$-symmetric systems \cite{Ozdmir2019}.
Because most photonic systems are usually lossy and gain is relatively difficult to implement, most previous works employed purely lossy systems with \textit{passive} $\mathcal{PT}$ symmetry for actual experiments.
The eigenfrequencies $\omega_{\pm}$ and eigenstates $v_{\pm}$ of the Hamiltonian in Eq. \ref{Hamiltonian} are
\begin{equation}
\label{EigenFreq}
    \omega_{\pm}
    =\omega_{0} +i \gamma_{0} \pm \sqrt{\kappa^{2} + (\Delta \omega +i \Delta \gamma)^{2}}
\end{equation}
and
\begin{equation}
\label{EigenState}
    v_{\pm}=
    \left( \begin{array}{c}
    1 \\ \frac{-(\Delta \omega +i \Delta \gamma) \pm \sqrt{\kappa^{2} + (\Delta \omega +i \Delta \gamma)^{2}}}{\kappa}
\end{array} \right),
\end{equation}
respectively. Here, the eigenstates are not normalized.
When the square root in Eq. \ref{EigenFreq} equals zero, there is only one solution for the eigenfrequency and eigenstate.
This is the condition for an EP.

Assuming that $\Delta\omega$ and $\kappa^{\prime\prime}$ are negligible, Eqs. \ref{EigenFreq} and \ref{EigenState} are reduced to the simple well-known equations
\begin{equation}
\label{EigenFreq2}
    \omega_{\pm}
    =\bar{\omega} +i \bar{\gamma} \pm \sqrt{\kappa^{\prime 2} - \Delta \gamma^{2}}
\end{equation}
and
\begin{equation}
    \label{EigenState2}
    v_{\pm}=
    \left( \begin{array}{c}
    1 \\ \frac{-i \Delta \gamma \pm \sqrt{\kappa^{\prime2} - \Delta \gamma^{2}}}{\kappa^{\prime}}
\end{array} \right).
\end{equation}
In the strong coupling limit where $\kappa^{\prime} >> \Delta\gamma$, the system behaves as a typical coupled system with two real eigenvalues and in-/anti-phase modes corresponding to $(1, \pm1)^{T}$ formed by mode hybridization.
This is called the $\mathcal{PT}$-symmetric phase.
In contrast, in the weak coupling limit where $\kappa^{\prime} << \Delta\gamma$, the eigenvalues become purely imaginary and the resonators oscillate independently as if there is no coupling between them.
Therefore, the corresponding eigenstates are $(1, 0)^{T}$ or $(0, 1)^{T}$, and the phase is called the $\mathcal{PT}$-broken phase.
Under the balanced condition, $\kappa^{\prime} = \Delta\gamma$, the eigenvalues and eigenstates coalesce, which means that only one mode can exist despite the coupling of the two resonator system. The singularity point is called  an EP in the NH system.    

From Eq. \ref{EigenState2}, when $\Delta\gamma>0$, the eigenstate at the EP can be written as
\begin{equation}
\label{EP}
    \left( \begin{array}{c}
    1 \\  -sign(\kappa^{\prime})i
    \end{array} \right).
\end{equation}
This implies that the relative phase difference between the two resonators is $\pi/2$ or $-\pi/2$, where the sign is determined by that of the coupling constant $\kappa^{\prime}$.
So far, only a few of the many reported studies on NH photonics have focused on phenomena at the EP \cite{Takata2017,Takata2021,Brandstetter2014}.
In particular, few studies have focused on the \textit{singular eigenstates} with a $\pm\pi/2$ phase difference expressed in Eq. \ref{EP} \cite{Lawrence2014}.
Moreover, the effects of the sign of the coupling constant $\kappa$ on the singular eigenstate at the EP have not been investigated in photonic systems because it is generally difficult to reverse the sign of the coupling in conventional photonic systems.

In this paper, we propose and numerically demonstrate a unidirectional radiation phenomenon manifested by Huygens dipoles formed at singular EP eigenstates in coupled plasmonic systems.
By exploiting the dipole-dipole interaction, two types of EPs due to negative and positive coupling and switching of the radiation direction from the Huygens dipoles are achieved.
This unique photonic functionality which directly originates from the singular eigenstates at the EPs provides new insights into NH photonic physics and devices.

\section{\label{sec:level2}Design and Principle}
\begin{figure}
\includegraphics[width=\linewidth]{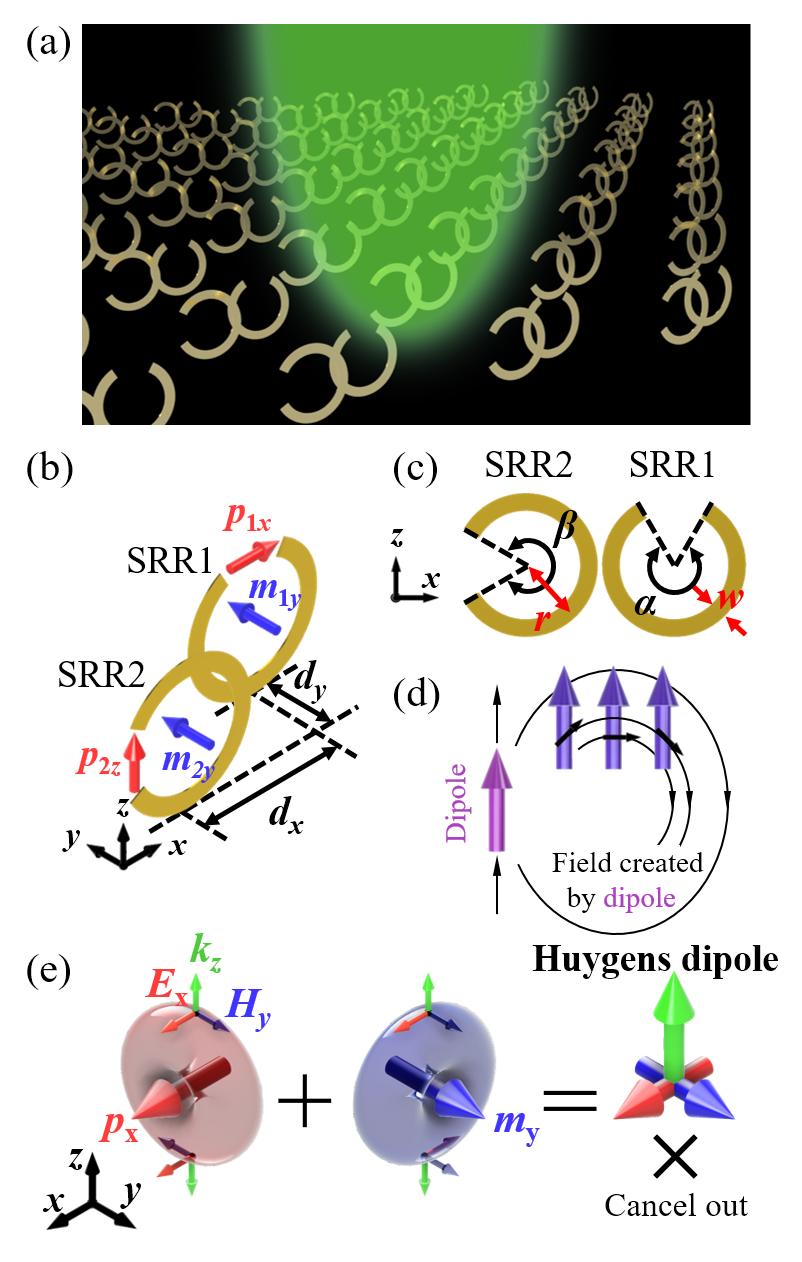}
\caption{\label{structure}(a)-(c) Schematics of the proposed coupled SRR pairs with different gap orientations aligned periodically along the $x$- and $y$- axes. $r$ = 6 mm, $w$ = 1.2 mm, $\beta$ = $300^\circ$, $d_{y}$ = 5 mm. $\alpha$ and $d_{x}$ are sweep parameters for finding the EPs. (d) A schematic of the dipole-dipole interaction. The sign of the coupling is determined by the direction of the force created by the other dipole. (e) Formation of Huygens dipoles composed of orthogonally arranged electric and magnetic dipoles with a relative phase difference of 0 or $\pi$.}
\end{figure}

A Huygens dipole is defined as a combination of an electric and a magnetic dipoles that are located at the same position and arranged orthogonally to each other, as shown in Fig. \ref{structure}(e) \cite{Picardi2018}.
Both the electric and magnetic dipoles radiate electromagnetic waves in the direction perpendicular to their dipole axes but they do so with different radiation phase profile: The phase profile is symmetric for electric dipoles, anti-symmetric for magnetic dipoles (Fig. \ref{structure}(e)).
As a result, only the radiation field in the direction of the cross-product of the two dipoles in a Huygens dipole survives because the radiation in the opposite direction is eliminated by destructive interference (Fig. \ref{structure}(e)).
Thus, to realize a Huygens dipole, orthogonally aligned electric and magnetic dipoles with a phase difference of 0 or $\pi$ are needed.

In this study, a C-shaped structure, which is known as a split ring resonator (SRR) in the metamaterials community, is employed to mimic electric and magnetic dipoles in an artificial structure.
The lowest eigenmode of the SRR is mainly composed of electric and magnetic dipole resonances \cite{Sersic2009}.
The electric dipole is parallel to the gap of the SRR, while the magnetic dipole is perpendicular to the plane of the SRR arc (Fig. \ref{structure}(b)).
We consider pairs of coupled SRRs arranged periodically in the $x$- and $y$- directions as shown in Fig. \ref{structure}(a).
The designed SRRs consist of perfect electric conductor (PEC), namely they do not have absorption loss.
The thickness of the SRRs is set to zero and the SRRs are assumed to be floating in air.
The two SRRs in each pair stand on the $x$-$y$ plane and are displaced from each other laterally in the $x$- and $y$-directions by $d_{x}$ and $d_{y}$.
While the electric and magnetic dipoles are located at the same position in a rigorous Huygens dipole, we realize the EP condition by introducing the finite shifts $d_{x}, d_{y}$.
The finite positional difference in the $x-y$ plane does not affect the directionality of the radiation, as discussed later.
For the detailed structural parameters, see the caption of Fig. \ref{structure}.

The different gap positions of the two SRRs result in different radiation loss rates.
A nonzero value of $\Delta \gamma$, implies that the coupled SRRs are described by a NH Hamiltonian with passive $\mathcal{PT}$ symmetry.
Because the SRRs are made of PEC, the dissipation of the resonances occurs only by radiation.
The radiation loss of the resonant modes is generally proportional to the amount of induced dipole moments that contribute to the far-field radiation.
In SRR1, as shown in Fig. \ref{structure}(b), both the electric dipole $\bm{p_{1x}}$ and the magnetic dipole \bm{$m_{1y}$} induced at the resonance can contribute to the far-field radiation in the $z$-direction because of the dipole orientation.
In contrast, in SRR2, the electric dipole $\bm{p_{2z}}$  cannot contribute to the z-directional radiation because of its dipole orientation, and only the magnetic dipole $\bm{m_{2y}}$ contributes to the far-field radiation.
Therefore, SRR1 has a larger amount of dipole moments and a larger radiation loss than SRR2 \cite{Moritake2017}.
In fact, 3D simulations for a single SRR reveal that the gap rotation can result in a reduction of $\gamma$ (Appendix A).

The coupling between the SRRs is mainly caused due to the dipole-dipole coupling of the induced magnetic dipoles because of the negligible electric dipole coupling due to the orthogonal arrangement of the SRRs and their sufficiently large spatial separation.
It is well known that the dipole-dipole interactions between two dipoles are determined by the relative orientations of the dipoles and the distance between them \cite{Arter2011,Liu2009}.
As shown in Fig. 1(d), when a dipole moves along the $x$-axis, the force experienced by the dipole varies from positive to negative depending on its position.
Because the sign of $\kappa^{\prime}$ in Eq. \ref{Hamiltonian} is opposite to that of the force, the strength and sign of $\kappa^{\prime}$ can be varied by increasing the $x-$direction shift $d_{x}$.
In general, negative coupling constants and sign reversal cannot be easily implemented in photonic systems such as coupled waveguides \cite{Guo2009,Ruter2010} and coupled ring resonators \cite{Peng2014}, although several complicated methods have been reported \cite{Keil2016}.
The ability to switch the sign easily is one key advantage of utilizing the dipole-dipole interaction.

We introduce a 2 $\times$ 2 Hamiltonian matrix to describe the proposed SRR system.
Writing the resonance amplitudes of SRR1 and SRR2 as $a_{1}$ and $a_{2}$, respectively, the Hamiltonian matrix can be written as
\begin{equation}
\label{Hamiltonian2}
\begin{split}
    \cal{H}
    \left(
    \begin{array}{c}
    a_{1} \\
    a_{2}
    \end{array} \right)
    &=
    \left( \begin{array}{cc}
    \Omega_{0} + \Delta \Omega & \kappa \\
    \kappa & \Omega_{0} -\Delta \Omega
    \end{array} \right)
    \left( \begin{array}{c}
    a_{1} \\
    a_{2}
    \end{array} \right),
\end{split}
\end{equation}
where $\Omega_{0}=\omega_{0}+i\gamma_{0}$ and $\Delta\Omega=\Delta\omega+i\Delta\gamma$.
As mentioned above, each SRR resonance is mainly composed of electric and magnetic dipole resonances.
An important point to note is that the induced magnetic dipoles are delayed in phase by $\pi/2$ compared to the induced electric dipole because the resonance originates from the LC resonance in the circuit model \cite{Sydoruk2009}.
In the LC resonance, there is a  phase difference of $\pi/2$ between the charge accumulation and current flow, which is similar to the relation between the induced electric and magnetic dipoles in SRRs.
Therefore, the total dipoles $\bm{\mu}_{i}$ induced in SRR1 and SRR2 can be written as 
\begin{equation}
\label{total}
    \bm{\mu}_{1}=\bm{p}_{1x}-ic\chi\bm{m}_{1y}, 
    \text{ }
    \bm{\mu}_{2}=\bm{p}_{2z}-ic\chi\bm{m}_{2y}, 
\end{equation}
where $c$ is the speed of light and $\chi=|m|/(|p|c)$ is a factor representing the amplitude ratio between the electric and magnetic dipoles induced in the SRRs at the lowest resonance.
The value of $\chi$ depends on the shape of the SRRs, as discussed later.
The factor of $-i$ on the right side of Eq. \ref{total} represents the phase delay of $\pi/2$.
When the EP condition is fulfilled, the eigenstate of Eq. \ref{Hamiltonian2} is $(a_{1}, a_{2})^{T}=(1, \pm i)^{T}$ from Eq. \ref{EP}.
Thus the total dipoles at the EP are
\begin{equation}
\label{TotalA}
\begin{split}
    \bm{\mu}_{1} \pm i \bm{\mu}_{2}
    &=\bm{p}_{1x} \pm c\chi\bm{m}_{2y} \pm i \left( \bm{p}_{2z} \mp c\chi\bm{m}_{1y} \right).\\
\end{split}
\end{equation}
If the Kerker condition \cite{Kerker1983},
\begin{equation}
\label{kerker}
    \chi=\frac{|\bm{m}|}{|\bm{p}|c}=1,
\end{equation}
is fulfilled, Eq. \ref{TotalA} can be written as the sum of Huygens dipoles, $\bm{h}=\bm{p}+c\bm{m}$, as follows:
\begin{equation}
\label{Huygens}
    \bm{\mu}_{1} \pm i \bm{\mu}_{2}=
    \begin{cases}
        \bm{h}_{z+}-i\bm{h}_{x-}   &   sign(\kappa)<0  \\
        \bm{h}_{z-}+i\bm{h}_{x+}   &   sign(\kappa)>0
    \end{cases}
\end{equation}
where $\bm{h}_{i\pm}$ is the Huygens dipole radiating in $\pm i$ direction.
Because light is emitted by $h_{x\pm}$ in the in-plane direction, only $h_{z\pm}$ contributes to the out-of-plane radiation.
Moreover, the sign of $\kappa^{\prime}$ determines the direction of the radiation from the Huygens dipole. 
This is the mechanism for the formation of Huygens dipoles and the switching of the radiation direction by using coupled SRRs under the EP condition.

Huygens dipoles have been utilized in many studies on Huygens metasurfaces \cite{Kim2014,Pfeiffer2013,Paniagua2015,Decker2015} and Huygens antennas \cite{Jin2010,Ziolkowski2015,Tang2016} in the metamaterials community.
In Huygens metasurfaces, induced electric and magnetic dipoles are elaborated by structural design to realize wavefront engineering and reflection-free directional propagation \cite{Kim2014,Pfeiffer2013,Paniagua2015,Decker2015}.
In these metasurface studies, the incident light from the outside is controlled, whereas in this study, the focus is on radiation from the eigenstates.
Therefore, the present study is more similar to Huygens antennas in terms of achieving directional radiation from the designed structure.
Huygens antennas are also composed of electric and magnetic elements to mimic Huygens dipoles and achieve directional radiation \cite{Jin2010,Ziolkowski2015,Tang2016}.
A Huygens antenna comprises a single pair of plasmonic elements, whereas the proposed structure is a periodic system.
The 0 or $\pi$ phase difference between the electric and magnetic dipoles in Huygens antennas is obtained by dual feed excitation \cite{Jin2010} or capacitive coupling \cite{Tang2016},  while the 0 or $\pi$ phase difference in this study is guaranteed by the EP condition in NH systems.
Therefore, NH physics plays an essential role in our approach.

\section{\label{sec:level3}Results and Discussions}
\begin{figure*}
\includegraphics[width=0.8\linewidth]{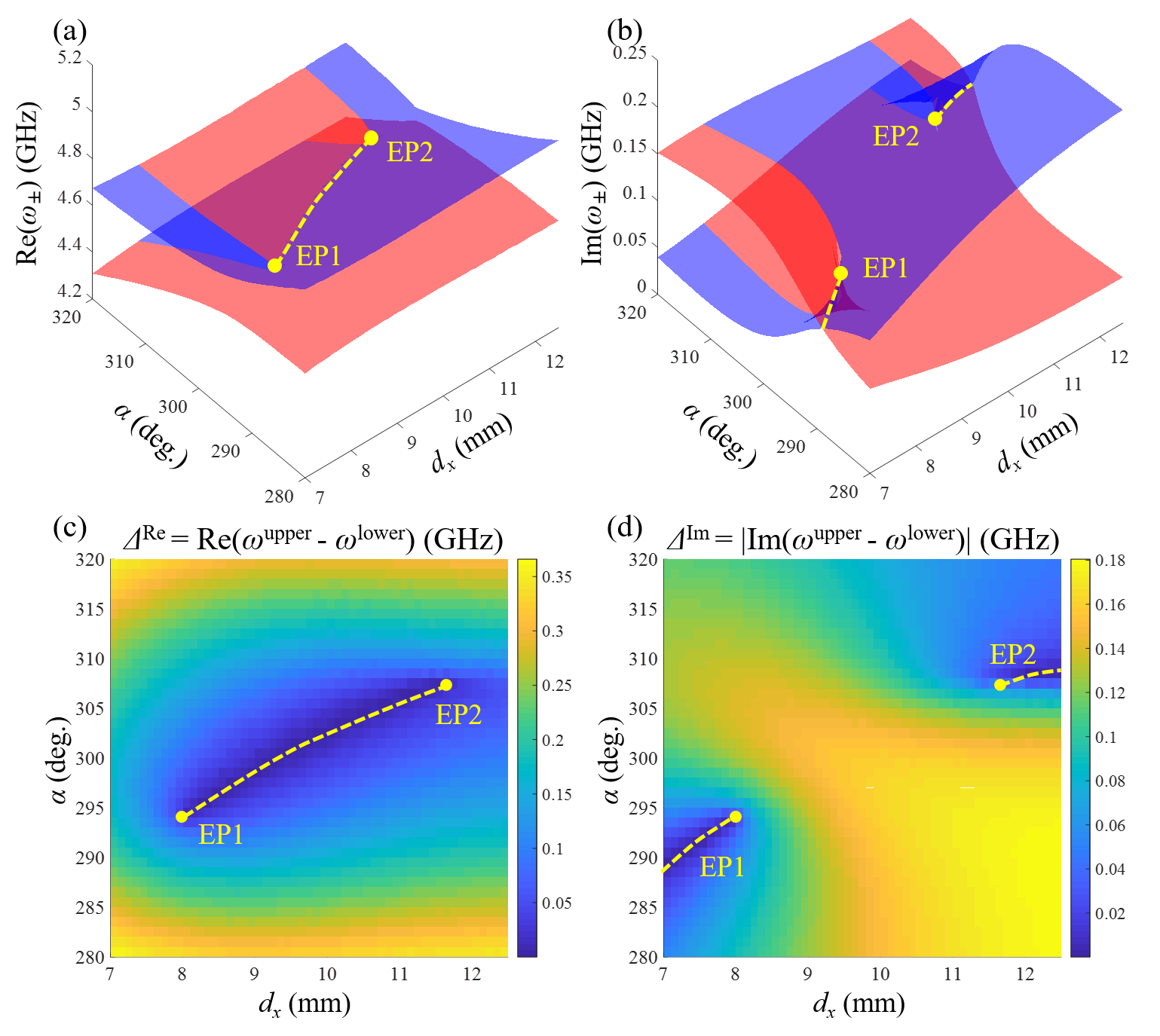}
\caption{\label{Riemann} (a) Real and (b) imaginary parts of the eigenfrequencies as functions of $d_{x}$ and $\alpha$. (c) Difference between real parts of upper and lower bands in (a). The blue valley line indicates the $\mathcal{PT}$-broken phase. (d) Absolute value of difference between imaginary parts of upper and lower bands in (b). The blue valley line indicates the $\mathcal{PT}$ symmetric phase.}
\end{figure*}

The simulated real and imaginary parts of the eigenfrequency as functions of $d_{x}$ and $\alpha$ are presented in Fig. \ref{Riemann}(a) and \ref{Riemann}(b), respectively.
The simulations were performed using the commercial finite element method solver (COMSOL).
Two EPs are observed, which are manifested as a self-intersecting Riemann surface structure in the complex eigenfrequencies.
The intersections of the Riemann surfaces at which the square root term in Eq. \ref{EigenFreq} is zero, are indicated by yellow dashed lines.
The sign of $\kappa^{\prime}$ varies from negative to positive as $d_{x}$ increases.
Therefore, the eigenstates $(a_{1} a_{2})^{T}$ at EP1 and EP2 correspond to $(1, i)^{T}$ and $(1, -i)^{T}$, respectively.

The real and imaginary parts of the frequency difference between the lower and upper bands are shown in Fig. \ref{Riemann}(c) and \ref{Riemann}(d), respectively.
The blue valley in Fig. \ref{Riemann}(c), which represents the $\mathcal{PT}$-broken phase, and connects the two EPs in the real eigenfrequency space.
In Fig. \ref{Riemann}(d), the valleys are extended from the EPs to the outside of the simulated parameter region, which represents the $\mathcal{PT}$-symmetric phase.
The eigenfrequencies are given by Eq. \ref{EigenFreq2} when the imaginary coupling $\kappa^{\prime\prime}$ in the system is absent.
It is then not necessary to tune $\Delta\omega$ and the yellow line in Fig. \ref{Riemann}(a) and \ref{Riemann}(b) (blue valleys in Fig. \ref{Riemann}(c) and \ref{Riemann}(d)) will be the constant line $\alpha\sim298^{\circ}$.
However, in the present system, not only  $\kappa^{\prime}$ and $\Delta\gamma$ but also $\kappa^{\prime\prime}$ and $\Delta\omega$ must be balanced to fulfill the EP condition in Eq. \ref{EigenFreq}.
$\kappa^{\prime}$ is controlled through $d_{x}$ to match $\Delta\gamma$, which is almost constant.
Moreover, because the system has finite imaginary coupling $\kappa^{\prime\prime}$, $\Delta\omega$ needs to be tuned through the arc angle of SRR1, $\alpha$, to eliminate $\kappa^{\prime\prime}$.
Therefore, $\alpha$ and $d_{x}$ need to be swept to determine the exact EPs.

\begin{figure}
\includegraphics[width=\linewidth]{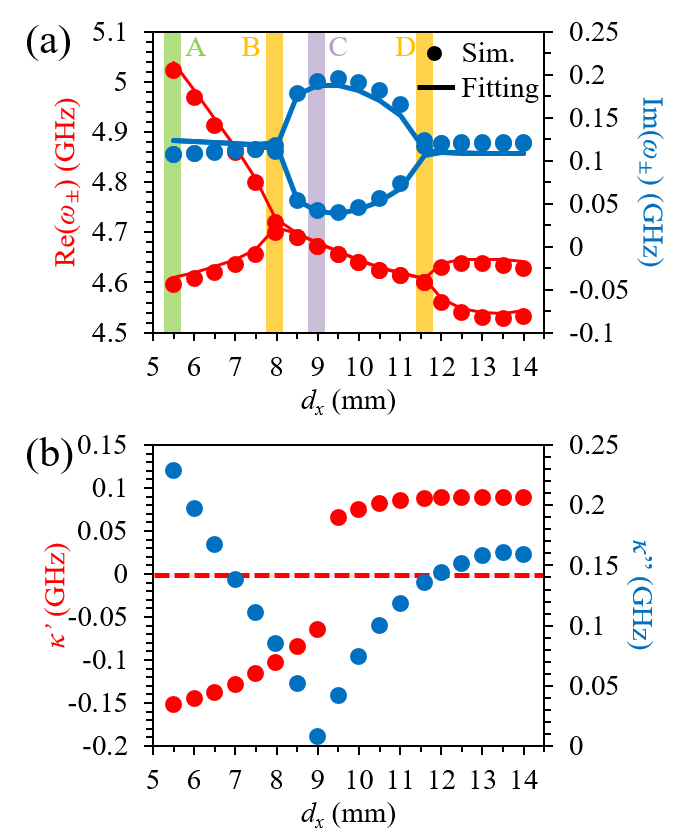}
\caption{\label{fitting} Results of eigenfrequency fitting along the yellow dashed line in Fig. \ref{Riemann}. (a) Simulated (circular data points) and fitted (lines) eigenfrequencies as functions of $d_{x}$. The red and blue data represent the real and imaginary parts of the eigenfrequencies, respectively. (b) Coupling constants obtained from the fitting. The red and blue data correspond to the real and imaginary parts, respectively.}
\end{figure}

The coupling constant $\kappa$ is generally difficult to calculate using numerical simulations although the parameters $\omega_{i}$ and $\gamma_{i}$ in Eq. \ref{Hamiltonian} can be directly calculated using a single-SRR model (Appendix A).
To determine the coupling constant from the calculated eigenfrequencies, fitting was carried out using Eq. \ref{EigenFreq}.
Figure \ref{fitting}(a) shows the results of fitting along the yellow dashed line in Fig. \ref{Riemann}.
The simulated data and fitted curves are indicated by circles and lines, respectively.
Two clear $\mathcal{PT}$ phase transitions, corresponding to the two EPs at around the $d_{x}$ values of approximately $8$ and $11.5$ mm, are observed.
As shown in Fig. \ref{fitting}(a), the simulated data are fitted well by Eq. \ref{EigenFreq}.
The averages of the differences between the simulated and the fitted values for the real and imaginary parts of the eigenfrequencies are $0.2\%$ and $7.6\%$, respectively.
In the fitting procedure, the $\omega_{i}$ and $\gamma_{i}$ calculated using the single-SRR model were set as constants (Appendix A).
The real and imaginary parts of the coupling $\kappa$ obtained by fitting are presented in Fig. \ref{fitting}(b).
Owing to the square form of $\kappa$ in Eq. \ref{EigenFreq}, the signs of $\kappa^{\prime}$ and $\kappa^{\prime\prime}$ are uncertain.
Here, we chose their signs so that the sign of the real part is flipped from negative to positive at around $d_{x} \sim 9$ mm.
Fig. \ref{fitting}(b) shows the finite values of $\kappa^{\prime\prime}$ determined in this manner.
Around $d_{x} \sim 9$ mm, where an abrupt change in $\kappa^{\prime}$ is observed, the real part should be zero if only the magnetic dipole interaction is present.
The finite value of $\kappa^{\prime}$ around $d_{x} \sim 9$ mm might be due to the presence of other coupling mechanisms such as electric dipole and higher-order multipole interactions \cite{Liu2009}. 

\begin{figure}
\includegraphics[width=0.7\linewidth]{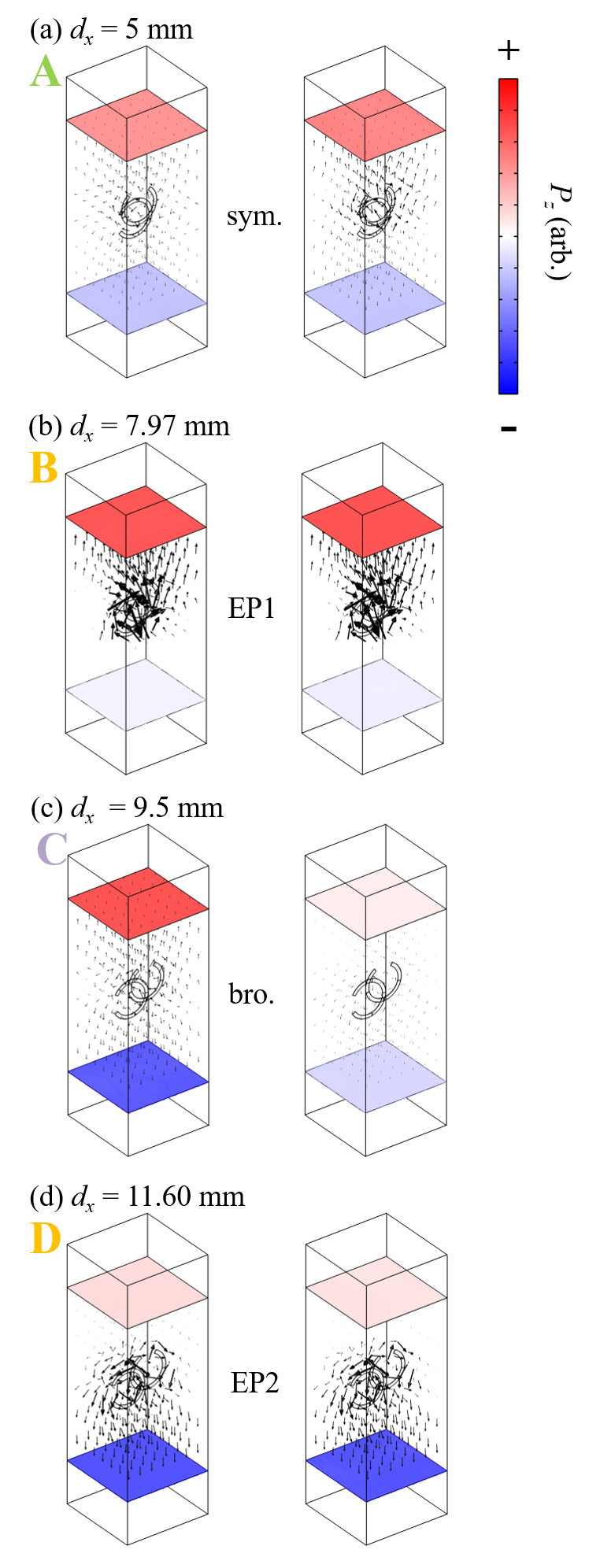}
\caption{\label{Pz} Poynting vector maps for $d_{x}$ = (a) 5, (b) 7.97, (c) 9.5, (d) 11.60, and (e) 14 mm. The $z$-components of the Poynting vectors are plotted at the upper and lower surfaces of the simulated model. The black arrows indicate the Poynting vector at each position.}
\end{figure}

As explained in the introduction, we expect directional radiation due to the formation of Huygens dipoles to be observed at the two EPs.
To confirm this, we plot the $z$-components of the Poynting vector $P_{z}$ at the upper and lower surfaces of the simulated models in Fig. \ref{Pz}.
The black arrows in Fig. \ref{Pz} indicate the direction and magnitude of the Poynting vector on a logarithmic scale.
The observation of a clear uni-directionality at EP1 and EP2 provides supporting evidence for the formation of Huygens dipoles.
Furthermore, the radiation direction reverses between EP1 and EP2.
Eq. \ref{Huygens} implies that this reversal corresponds to a flip in the sign of $\kappa^{\prime}$ from negative to positive with the increase in $d_{x}$.
Under the EP conditions (Figs. \ref{Pz}(b) and \ref{Pz}(c)), the Poynting vectors are aligned along the $z$-direction and the radiation in the in-plane direction is negligible far away from the SRRs, while the directions of the Poynting vectors show complex behavior in the vicinity of the SRRs. 
Here, two plots are shown at each EP although the eigenstates should, in theory, coalesce at the EP.
However, in the actual simulations, the finite numerical errors result in two very similar eigenfrequencies and eigenstates, and we present both plots to show that the similar eigenstates are obtained.
In contrast to the radiation patterns at the EPs, the radiation patterns are symmetric along the $z$ direction in the $\mathcal{PT}$-symmetric and $\mathcal{PT}$-broken phases.
This implies that the unidirectionality of the radiation is caused by the singular eigenstates at the EPs.

\begin{figure}
\includegraphics[width=\linewidth]{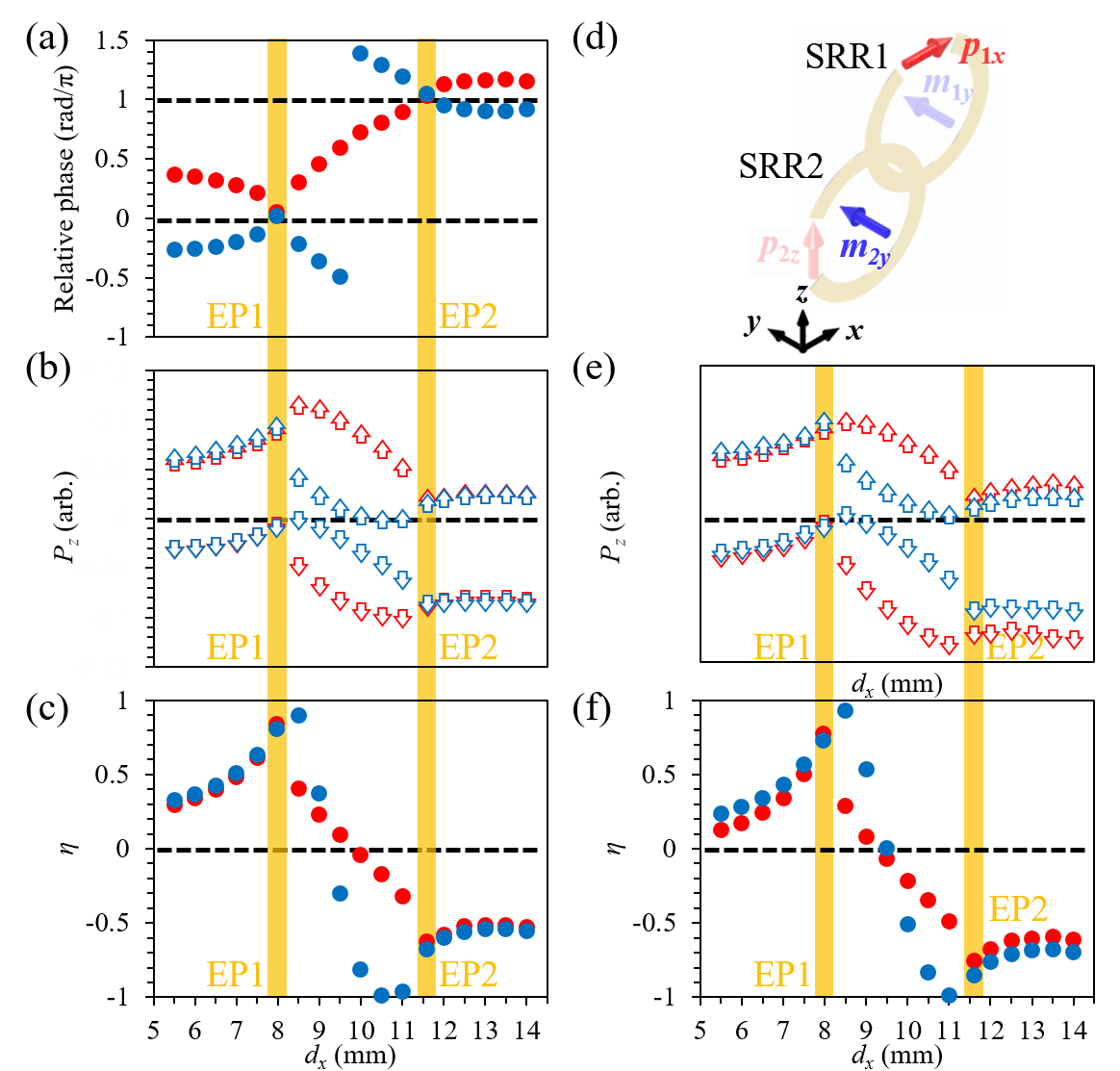}
\caption{\label{directional} Relative phase and quantitative evaluation of directionality. The red (blue) data correspond to the results from eigenfrequencies with a smaller (larger) real part in the $\mathcal{PT}$-symmetric phase and a smaller (larger) imaginary part in the $\mathcal{PT}$-broken phase. (a) Relative phase between $\bm{p}_{1x}$ and $\bm{m}_{2y}$ obtained from the simulated field profiles. (b) The integral of $P_{z}$ in each plane, $I_{\pm}$ and (c) $\eta$ defined in Eq. \ref{directionality} for different $d_{x}$. (d) The dipole model composed of electric and magnetic dipoles. (e) $I_{\pm}$ and (f) $\eta$ calculated for the dipole model at different $d_{x}$.}
\end{figure}

To confirm the formation of Huygens dipoles, the relative phases between $\bm{p_{1x}}$ and $\bm{m_{2y}}$ obtained from the simulated field profiles are shown in Fig. \ref{directional}(a).
As expected, the phase differences are 0 and $\pi$ at EP1 and EP2, respectively, which is exactly the condition for a Huygens dipole.
To evaluate the directionality quantitatively, the integral of $P_{z}$ in each plane in Fig. \ref{Pz}, $I_{\pm}$, is shown in Fig. \ref{directional}(b). 
We define a parameter $\eta$ to quantify the degree of directionality
\begin{equation}
\label{directionality}
    \eta = \frac{|I_{+}|-|I_{-}|}{|I_{+}|+|I_{-}|},
\end{equation}
and plot it in Fig. \ref{directional}(c).
When $\eta$ equals 1 (-1), all the radiation is emitted to the upper (lower) side of the system.
The red (blue) data correspond to the results from eigenfrequencies with a smaller (larger) real part in the $\mathcal{PT}$-symmetric phase and a smaller (larger) imaginary part in the $\mathcal{PT}$-broken phase.
As shown in Fig. \ref{directional}(c), $\eta$ increases in the vicinity of the EPs and has clearly opposite signs near the two EPs each other.
Some of the blue points in the $\mathcal{PT}$-broken phase have larger values of $\eta$ than those at the EPs because of the phase retardation caused by a finite vertical ($z$-direction) deviation between the electric and magnetic dipoles, as discussed later.
Although these blue points have large $\eta$ values, the radiation itself is weak because of the small radiation loss (small imaginary part), as shown in Fig. \ref{directional}(b).
From the analysis of the relative phase and directivity, we conclude that the unidirectional emission is due to the formation of the Huygens dipole at the EP and the switching of the unidirectional emission is due to the sign reversal of the real part of the coupling constant. 

In the last part of this paper, we show that the proposed system and observed phenomena can be understood using a simple dipole model.
The dipole model is composed of electric and magnetic dipoles as shown in Fig. \ref{directional}(d).
The electric dipole is placed at the center position of the gap of SRR1, and the magnetic dipole is arranged at the center of the arc of SRR2.
The model simulations were carried out using the same method as the previous simulations (COMSOL), except that the point dipole sources were used instead of the SRRs.
The frequency, relative amplitude, and relative phase between the SRRs obtained from simulations were used in the model simulations.
The $I_{\pm}$ and $\eta$ defined in Eq. \ref{directionality} calculated using the dipole model shown in Fig. \ref{directional}(e) and (f) are in good agreement with those calculated in the simulations shown in Fig. \ref{directional}(b) and \ref{directional}(c).
This indicates that the coupled SRRs can be modeled by electric and magnetic dipoles and that the uni-directionality can be understood by the formation of the Huygens dipole.

Here, we consider the discrepancies in the amplitude ratio $\chi$ and the vertical deviation of the electric and magnetic dipoles between a Huygens dipole formed in coupled SRRs at the EPs from those of a rigorous Huygens dipole.
To form a rigorous Huygens dipole, not only it is necessary for the relative phase to be $0$ or $\pi$, but the Kerker condition in Eq. \ref{kerker} (i.e., $\chi=1$) must also be satisfied.
The relative amplitude between the electric and magnetic dipoles of the lowest mode in SRRs generally depends on the shape of the SRRs and can only be obtained through an analysis of the spectra \cite{Husnik2008}.
In the model simulations, we varied the ratio $ \chi =|\bm{m}|/(|\bm{p}|c)$.
The results at $\chi=0.7$ are presented in Fig. \ref{directional}(e) and (f) because these results seem to show better agreement with the simulations in Fig. \ref{directional}(b) and \ref{directional}(b).
The value of $\chi=0.7$ is consistent with the value roughly estimated from the imaginary part of the eigenfrequencies in a single SRR (Appendix B).
In the designed coupled SRRs, the induced electric and magnetic dipoles that form the Huygens dipole are at different positions, whereas the rigorous Huygens dipole is defined by the sum of dipoles at the same position.
The induced electric and magnetic dipoles are shifted in the $x-y$ plane by $d_{x}$ and $d_{y}$ and there are finite vertical deviations between the dipoles because the induced electric and magnetic dipoles are located at the gap and center of the SRRs, respectively. 
By investigating the effect of the positional difference on the directionality (Appendix B), it was found that displacements along the radiation ($z$-) direction strongly affect $\eta$ while lateral displacements in the $x$-$y$ plane do not have a significant effect.
The effect of the vertical deviation can be reduced by introducing $z-$ directional shift between the SRR1 and SRR2.
The achieved directivity of approximately $0.8$ at the EPs is sufficient to demonstrate the unidirectionality of the radiation by the Huygens dipole although the highest directivity it still obtained at the optimal conditions of $\chi=1$ and zero shift in $z$-direction.

\section{\label{sec:level4}Summary}
We have theoretically and numerically investigated coupled SRRs described by a NH Hamiltonian in which non-Hermiticity is induced by differences in radiation losses. 
We demonstrated the formation mechanism of a Huygens dipole under EP conditions by employing electric and magnetic dipole components in the lowest mode of the coupled SRRs.
The the two EPs are induced by negative and positive coupling constants realized through dipole-dipole interaction.
Uni-directional radiation originating from the singular eigenstates at the EPs and the reversal of the radiation direction by flipping the signs of the coupling constants were demonstrated.
We showed that the unidirectionality of the EPs in the SRRs can be reproduced by the simple dipole model.
To date, singular eigenstates at the EPs have been investigated using the Jones matrix for circular polarization \cite{Lawrence2014}, and the scattering matrix for unidirectional phenomena \cite{Feng2013}.
The utilization of singular EP eigenstates in the \textit{Hamiltonian matrix} in this study provides a new perspective for the use of such states in photonic devices.

\begin{acknowledgments}
This work was supported 
by JST, PRESTO Grant Number JPMJPR18L9, Japan,
by JSPS, Grant-in-Aid for Scientific Research (S), 20H05641, Japan,
and by Grant-in-Aid for Early-Career Scientists, 21K14551.
\end{acknowledgments}

\appendix
\renewcommand{\thefigure}{A\arabic{figure}}
\setcounter{figure}{0}
\section{Eigenfrequecy of single SRR}
\begin{figure}
\includegraphics[width=\linewidth]{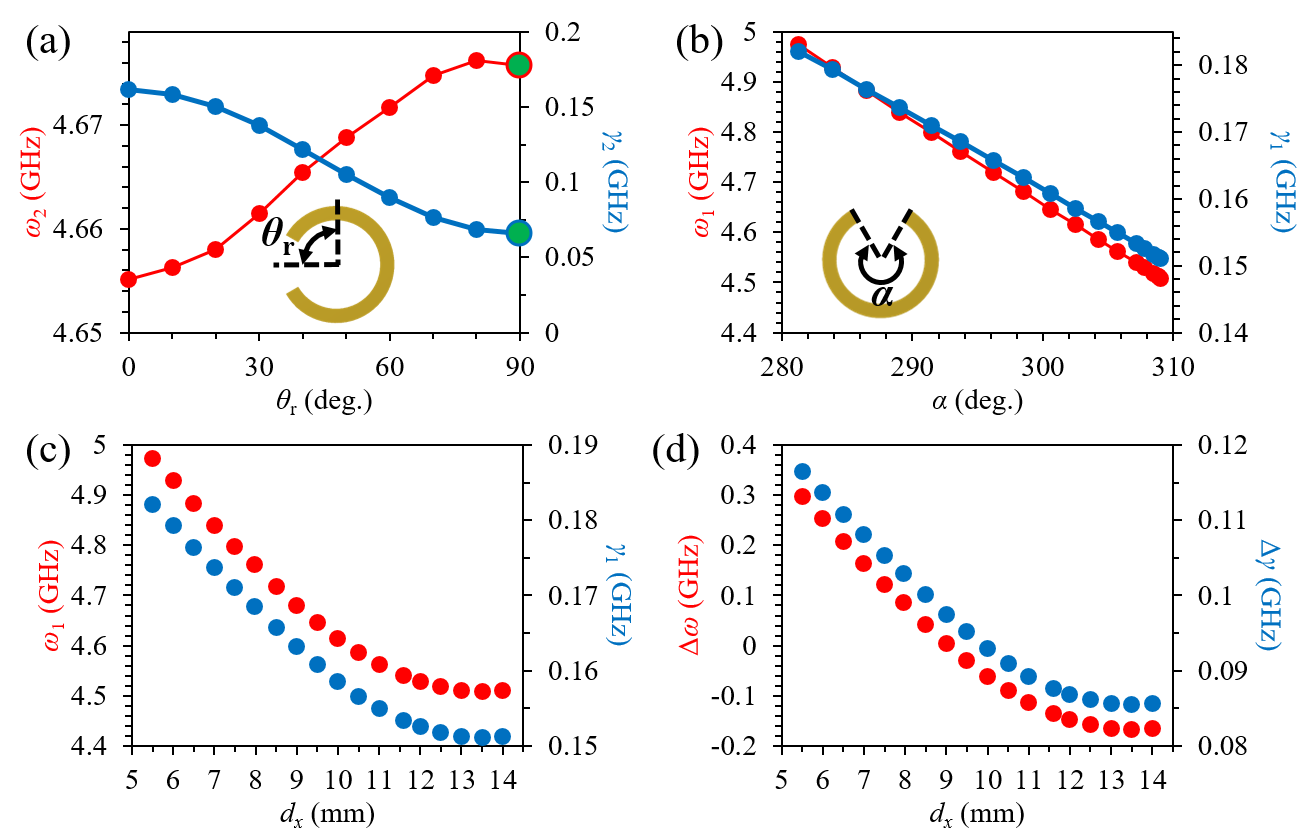}
\caption{\label{single} Real and imaginary parts of eigenfrequencies of a single SRR as functions of (a) the rotation angle $\theta_{r}$, (b) the arc angle $\alpha$, and (c) $d_{x}$. (d) $\Delta\omega$ and $\Delta\gamma$ calculated from (a) and (c).}
\end{figure}

The simulated eigenfrequencies for a single SRR are presented in Fig. \ref{single}.
Figure \ref{single}(a) shows the real and imaginary parts of the eigenfrequency as functions of the rotation angle $\theta_{r}$.
As mentioned in the main text, the imaginary part is reduced with the rotation of the gap position because of the decreased radiation loss.
The green dots denote the values used for fitting $\omega_{2}$ and $\gamma_{2}$ in the main text.
Figure \ref{single}(b) shows the real and imaginary parts of the eigenfrequency as functions of the arc angle $\alpha$.
A longer arc length leads to a lower resonant frequency.
The imaginary part also has the same variation with $\alpha$.
Figure \ref{single}(c) shows the values of $\omega_{1}$ and $\gamma_{1}$ used in the fitting as functions of $d_{x}$.
Figure \ref{single}(d) shows $\Delta\omega$ and $\Delta\gamma$ calculated from Figs. \ref{single}(a) and \ref{single}(c).

\renewcommand{\thefigure}{B\arabic{figure}}
\setcounter{figure}{0}
\begin{figure}
\includegraphics[width=\linewidth]{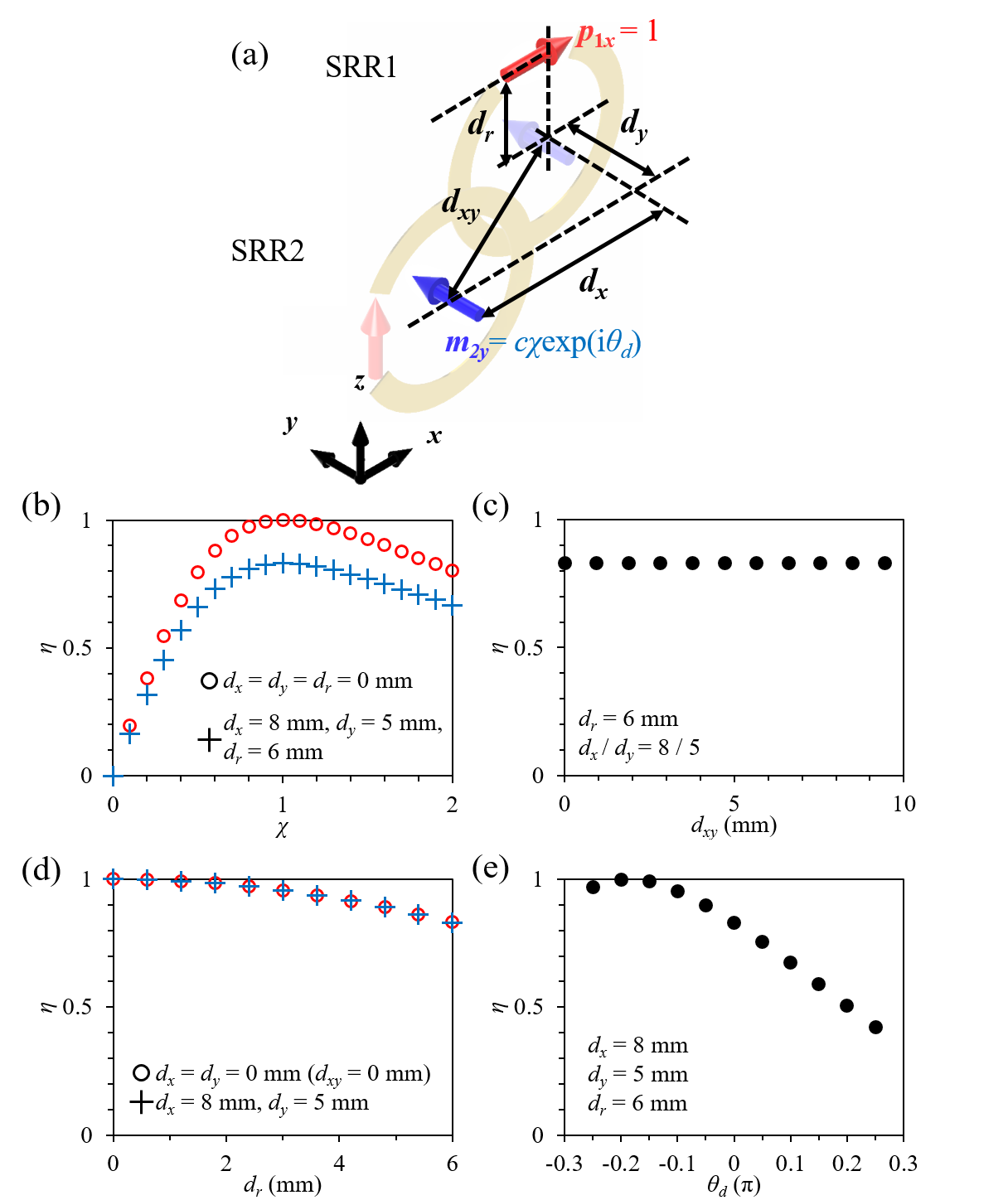}
\caption{\label{model} (a) A schematic of the electric and magnetic dipoles formed in SRRs for model simulations. (b) Directionality $\eta$ as a function of $\chi$ for rigorous Huygens dipoles (red circles) and proposed SRRs (blue crosss). (c) $\eta$ as a function of $d_{xy}$ with $d_{x}/d_{y}=8/5$. (d) $\eta$ as a function of $d_{r}$ (e) $\eta$ as a function of $\theta_{d}$.}
\end{figure}
\section{The effect of discrepancy from rigorous Huygens dipole}
Figure \ref{model} shows the effect of the discrepancy between the rigorous Huygens dipole and the Huygens dipole composed of induced electric and magnetic dipoles in the SRRs on $\eta$.
The relative positions of the dipole sources are shown in Fig. \ref{model}(a).
$d_{xy}$ is the distance between $\bm{p_{1x}}$ and $\bm{m_{2y}}$ in the $x,y$-plane, and $d_{r}$ is the distance along the $z$-axis.
The amplitude ratio and phase difference are $c\chi$ and $\theta_{d}$, respectively.
The values of $\chi$ was set to 1 and $\theta_{d}$ to $\pi/2$ for the model simulations described hereafter.

The effect of the amplitude ratio is first investigated.
Figure \ref{model} (b) shows the directionality $\eta$ as a function of $\chi$.
The circles and crosses denote the results for the rigorous Huygens dipole ($d_{x}=d_{y}=d_{r}= 0$ mm) and the proposed SRRs ($d_{x}=8$ mm, $d_{y}=5$ mm, $d_{r}= r = 6$ mm), respectively.
For both cases, $\eta$ reaches its highest value at $\chi=1$, which is the Kerker condition in Eq. \ref{kerker}.
In the proposed structure, $\eta$ reaches up to 0.8 despite its degradation by the positional difference between the dipoles.
Therefore, the amplitude ratio $\chi=0.7$ is sufficient to obtain a high radiation directivity.

In Ref. \cite{Zeng2010,Sersic2011}, the ratio $\chi$ in the lowest resonance of the SRR was estimated from the scattering cross-section of the SRRs.
Generally, the scattering cross-section is proportional to the radiation loss, that is the imaginary part of the eigenfrequencies of the SRRs.
Here, we estimate the ratio $\chi$ from the imaginary parts of the eigenfrequencies of a single SRR.
Fig. \ref{single}(a) shows that the dipole contributions to the imaginary part can be controlled  by rotating the SRR.
At $\theta_{r}=0^{\circ}$, both the electric and magnetic dipoles contribute to the radiation while only the magnetic dipole contributes at $\theta_{r}=90^{\circ}$.
Therefore, the imaginary part $\gamma(\theta_{r})$ can be divided into contributions from the electric dipole $\gamma_{ED}$ and magnetic dipole $\gamma_{MD}$ as follows:
\begin{equation}
\begin{split}
    \gamma(\theta_{r}=0^\circ)&=\gamma_{ED}+\gamma_{MD},\\
    \gamma(\theta_{r}=90^\circ)&=\gamma_{MD},
\end{split}
\end{equation}
The ratio $\chi$ can then be calculated as
\begin{equation}
\label{ratio}
    \chi=\frac{\gamma_{MD}}{\gamma_{ED}}=\frac{\gamma(\theta_{r}=90^\circ)}{\gamma(\theta_{r}=0^\circ)-\gamma(\theta_{r}=90^\circ)}.
\end{equation}
$\chi$ was calculated to be 0.69 at $\alpha=300^\circ$, which is consistent with the values used for the directivity model simulations in the main text.

Next, the effect of the positional difference of the dipoles is investigated.
Figure \ref{model}(c) shows $\eta$ as a function of $d_{xy}$ while $d_{x}/d_{y}$ is maintained at $8/5$.
Figure \ref{model}(c) shows that $\eta$ is almost constant, which implies that the shift in the $x,y$-plane does not affect $\eta$.
In contrast, $\eta$ approaches 1 as $d_{r}$ decreases, as shown in Figure \ref{model}(d).
Therefore, $\eta$ is degraded by the spatial deviation along the radiation direction because of the phase retardation between the fields radiated by the dipoles.
The phase retardation can be compensated by changing the phase difference $\theta_{d}$, as shown in Fig. \ref{model}(e).
$\eta$ reaches 1 at around $\theta_{d}=-0.2\pi$, which is consistent with the high values of $\eta$ in the blue data in Fig. \ref{directional}(c) and \ref{directional}(f).
\bibliography{moritake}

\end{document}